\documentclass[twocolumn,trackchanges]{aastex701}

\usepackage{mathtools, amsmath}
\usepackage{nccmath}
\usepackage{float}
\usepackage{graphicx}
\usepackage{fontawesome}
\usepackage{hyperref}
\usepackage{placeins}
\usepackage{ulem}
\usepackage{multirow}
\usepackage{hhline}
\usepackage{xcolor}
\usepackage{tabularx,booktabs}
\usepackage[utf8]{inputenc}
\usepackage[T1]{fontenc}
\usepackage{textcomp}   
\defcitealias{Hosenie_2021}{MeerLICHT}
\defcitealias{Groot_2024}{BlackGEM}
\defcitealias{pessto}{ePESSTO+}
\defcitealias{Hodgkin_2021}{GaiaAlerts}

\begin{document}

\title{The Weighing Halos Accurately, Locally, and Efficiently with Supernovae (WHALES) Survey Overview and Initial Data Release}

\author[0000-0002-5389-7961]{Maria Acevedo}
\affiliation{Department of Physics, Duke University, Durham, NC 27708, USA}
\email[show]{ma474@duke.edu}  

\author[0000-0002-4934-5849]{Daniel Scolnic}
\affiliation{Department of Physics, Duke University, Durham, NC 27708, USA}
\email{daniel.scolnic@duke.edu}

\author[0000-0002-7234-844X]{Bastien Carreres}
\affiliation{Department of Physics, Duke University, Durham, NC 27708, USA}
\email{bastien.carreres@duke.edu}

\author[0000-0001-8596-4746]{Erik R.~Peterson}
\affiliation{Department of Physics, Duke University, Durham, NC 27708, USA}
\affiliation{Department of Physics, University of Michigan, Ann Arbor, MI 48109, USA}
\email{erik.r.peterson@duke.edu}

\author[0000-0002-8687-0669]{Bruno O. Sánchez}
\affiliation{Aix Marseille Univ, CNRS/IN2P3, CPPM, Marseille, France}
\email{bsanchez@cppm.in2p3.fr}

\author[0000-0003-1731-0497]{Christopher Lidman}
\affiliation{The Research School of Astronomy and Astrophysics, The Australian National University, Canberra, ACT 2611, Australia}
\email{Christopher.Lidman@anu.edu.au}

\author[0009-0006-4963-3206]{Bailey Martin}
\affiliation{The Research School of Astronomy and Astrophysics, The Australian National University, Canberra, ACT 2611, Australia}
\email{Bailey.Martin@anu.edu.au}

\author[0000-0003-0017-349X]{Christopher A. Onken}
\affiliation{The Research School of Astronomy and Astrophysics, The Australian National University, Canberra, ACT 2611, Australia}
\email{christopher.onken@anu.edu.au}

\author[0000-0002-6124-1196]{Adam G.~Riess}
\affiliation{Space Telescope Science Institute, Baltimore, MD 21218, USA}
\affiliation{Department of Physics and Astronomy, Johns Hopkins University, Baltimore, MD 21218, USA}
\email{ariess@stsci.edu}

\begin{abstract}

We present an overview of the Weighing Halos Accurately, Locally, and Efficiently with Supernovae (WHALES) survey, the first to discover and measure Type Ia supernovae (SNe~Ia) in and around galaxy superclusters. By building a sample of SNe~Ia around these massive environments, we aim to provide new constraints on bulk-flow models while laying the groundwork for improved estimates of supercluster masses. Here, we present data from the first two seasons targeting the Shapley Supercluster ($0.02<z<0.06$), which is responsible for a large but unknown fraction of our local group's motion. Until now, no supernovae had been analyzed in the direction of Shapley. Through the WHALES survey, we have identified 12 likely SNe~Ia in this region using SkyMapper, including 8 with spectroscopic confirmation. We present the first light curves of these SNe and combine our observations with data from ATLAS. We demonstrate that the low number of discovered SNe Ia per season is consistent with various rate calculations, highlighting the need for future surveys to monitor superclusters over a multi-year timespan. Finally, we present simulations of SN~Ia observations in the environments of massive galaxy clusters, demonstrating how the inferred peculiar velocities can constrain cluster masses, and highlighting the added complexity within superclusters. We find that a sample of 100 SNe~Ia would enable a 25\% precision measurement of the total mass of the Shapley Supercluster.

\end{abstract}



\section{Introduction} 
Galaxy redshift surveys and Type Ia supernovae (SNe~Ia) are complementary tools for mapping the matter distribution and modeling peculiar velocities \citep{Scolnic_2018, Tully_2019, Brout_2019, Peterson_2022}. A central question is the origin of the Milky Way’s $\sim$600~km/s motion relative to the cosmic microwave background (CMB) \citep{Tully_2019}. Although this bulk flow is well-measured, its physical drivers are not fully understood. Current reconstructions suggest that roughly half of the motion can be attributed to nearby clusters such as the nearer Laniakea Supercluster and the more distant Shapley Supercluster, with the remainder arising from a large underdensity known as the Local Void, or ``Dipole Repeller’’ \citep{Tully_2019, Tully_2023}. Furthermore, some measurements of bulk flows on larger scales indicate potential tensions with the $\Lambda$CDM model of cosmology \citep{whitford2023}. Progress in resolving these discrepancies has been limited by the fact that most peculiar velocity reconstructions do not extend beyond the Shapley Supercluster ($z~\ge$~0.06) due to the limited redshift depth of nearby galaxy surveys \citep{Lavaux_2011}. This gap in coverage makes it difficult to directly assess Shapley’s contribution to the local velocity field.

Understanding the gravitational pull of the Shapley Supercluster is crucial for correcting peculiar velocities in cosmological analyses with SNe~Ia \citep{Scolnic_2018, Brout_2019, Peterson_2022}. While \cite{Peterson_2022} estimate that uncertainties from peculiar velocity models contribute roughly $\sim$0.2 km/s/Mpc to the errors in the Hubble constant $H_0$, they show that the impact of these corrections can be up to three times higher. Moreover, \cite{union} estimate that peculiar-velocity correlations represent the largest systematic uncertainty in measurements of the evolution of the dark energy equation-of-state parameter, $w_a$. Here, we attempt to discover and measure SNe~Ia in the direction of the Shapley Supercluster as a unique opportunity to validate current models (e.g. \cite{cf4}).

The Shapley Supercluster is one of the most massive and dense structures in the nearby universe, spanning roughly 1800 Mpc $\times$ 3600 Mpc ($\sim 200$ square degrees), a redshift range of $0.02 < z < 0.06$, and hosting numerous rich clusters \citep{stopyra2021}. Its dominant subcluster, A3571, has mass estimates that vary significantly between dynamical, X-ray, and Sunyaev–Zel’dovich methods, highlighting the challenge of accurately characterizing even a single subcomponent of this extreme environment \citep{stopyra2021}. Superclusters like Shapley are rare, and the number of such massive clusters within the local super-volume varies substantially depending on the mass estimation method, making them an important test of $\Lambda$CDM cosmology.

Currently, of the $\sim$650 low-redshift SNe~Ia used for peculiar velocity studies, only six are within 300 square degrees of the Shapley Supercluster \citep{Scolnic_2022}. This scarcity is largely due to observational constraints—Shapley lies near a declination of $-30^{\circ}$, a region that can only be observed from the southern hemisphere. Additionally, the amount of Milky Way (MW) dust extinction in this region of the sky is relatively high (MW$E(B-V) = 0.15$~mag), and surveys typically avoid areas with higher extinction to maximize the depth of their observations.

Existing low-redshift surveys provide limited coverage of this region. The Dark Energy Bedrock All-Sky Supernovae (DEBASS) program observes low-z SNe with DECam, but primarily for follow-up, and has only 77 publicly available SNe to date \citep{nora}. The Shapley region also lies near the southern boundary of the Zwicky Transient Facility footprint (ZTF; \citealp{ztfdr2}) and within the footprint of the Young Supernova Experiment (YSE; \citealp{Jones2021, Chambers16}), leaving it poorly sampled overall.

In this analysis, we present a dedicated search and follow-up program for SNe~Ia in the direction of the Shapley Supercluster. In Section~\ref{sec:survdet}, we discuss our survey design, in Section~\ref{sec: datared}, we discuss how our images are processed and calibrated, and how we build our simulations, and in Section~\ref{sec:datahubble}, we show light curves for our sample as well as a preliminary Hubble diagram. Discussions and conclusions are in Sections~\ref{sec:disc} and~\ref{sec:concl}.

\section{Survey Overview} \label{sec:survdet}
Weighing Halos Accurately, Locally, and Efficiently with Supernovae (WHALES) is a new discovery survey designed to increase the number of supernovae around the Shapley Supercluster. We have observed for two seasons thus far: Season 1, from January 2023 to August 2023, and Season 2, from January 2024 to August 2024. 

\subsection{Instrumentation Overview} \label{sec:instov}
Images for WHALES were obtained using the SkyMapper telescope at Siding Spring Observatory in Australia. SkyMapper is a 1.35 meter, reflecting, modified Cassegrain telescope with an approximately 5 square degree field-of-view \footnote{During our survey, only half of the CCD mosaic was operational, giving a field of view of 2.6 square degrees per image.} and a pixel size of 0.5\arcsec \citep{smdr4}. Each CCD comprises 2048$\times$4096 pixels, yielding $\sim$8$\times$10$^7$ pixels per pointing across the array. WHALES observations use the SkyMapper $g$ (4000–5600 \AA), $r$ (5600–7000 \AA), and $i$ (7000–8200 \AA) bands \citep{smfilt} which are comparable to The Legacy Survey of Space and Time (LSST) filters \citep{lsst, lsst_filters}. For this survey, we obtained time on SkyMapper corresponding to approximately three hours of useful observing time per week during Season 1 and six hours per week during Season 2, on average, over a six-month observing window each season. WHALES observed for 60 nights in Season 1 and 70 nights in Season 2.

\subsection{Observing Strategy} \label{sec:obvsstrat}
\begin{figure}
    \centering
    \includegraphics[width=\columnwidth]{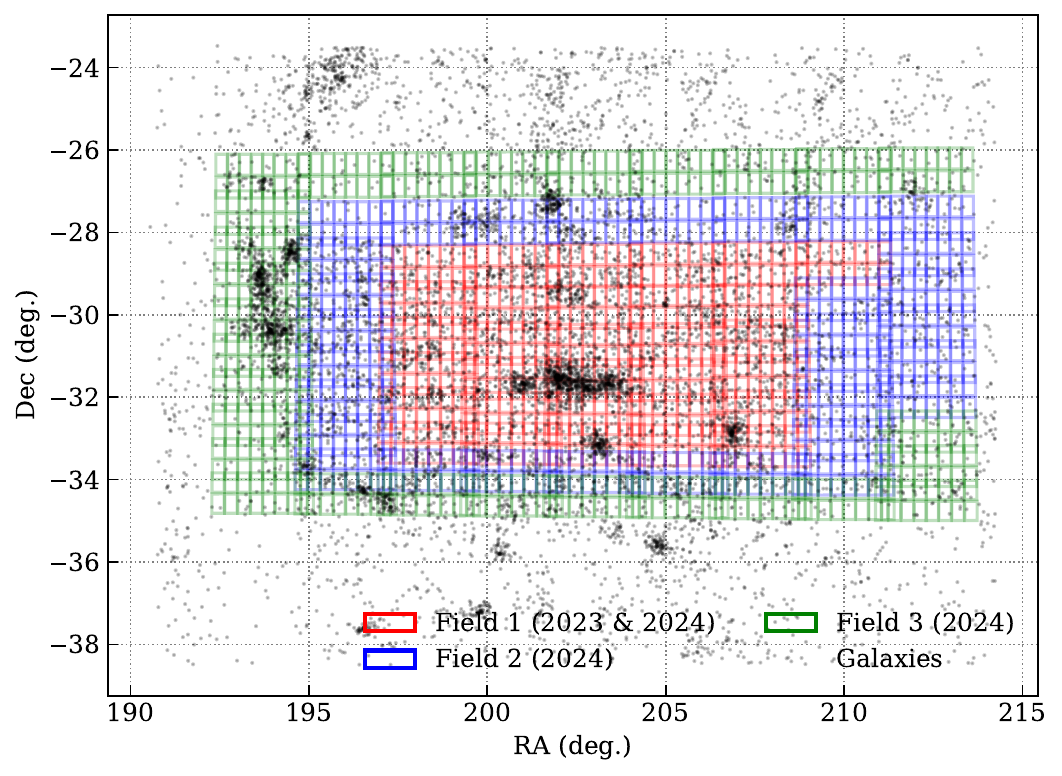}
    \caption{A visualization of our survey design. The black points are galaxies located in or near the Shapley supercluster as identified in \citet{Quintana_2020}. The boxes represent the CCD layout of SkyMapper, which is in a $2\times8$ grid. The red boxes are Field 1 (Season 1 \& 2), while the blue and green boxes are Fields 2 and 3 (Season 2), respectively.}
    \label{fig:obsplan}
\end{figure}
The WHALES survey is designed to monitor the full extent of the Shapley Supercluster, with a tiling plan that begins at the dense central region and extends outward. This strategy accounts for instrument availability, observing season, and galaxy density using the \citet{Quintana_2020} redshift catalog, with observations prioritized in the cluster core where galaxy populations and radial motions are strongest (Figure~\ref{fig:obsplan}). In Season 1, we implemented a $5\times6$ grid of pointings centered on $(RA, DEC) = (13^{h}25^{m}00^{s}, -31^{\circ}00'00'')$, covering $\sim$30\% of the cluster each night with 30 unique pointings per filter (Field 1). Exposures were 100 seconds, and with an average of 22 seconds overhead between pointings, we obtained $\sim$3 hours of observations per night. Typical conditions yielded 3.34\arcsec\ seeing and a $g$-band depth of 18.7 mag. In Season 2, we expanded coverage to the full 200 deg$^2$ supercluster by adding two fields in a concentric spiraling slewing pattern (Field 2 with 26 pointings and Field 3 with 29 pointings), tiling the area over three nights. The Season 2 observations maintained 30 pointings in each of the three bands each night, with an average seeing of 3.18\arcsec and a $g$-band depth of 18.5 mag.

Our goal was to observe each field with a 3-5 day cadence (similar to ZTF as described in \citealp{ztfdr2}) to ensure comprehensive pre-peak and post-peak data for any potential SNe~Ia. However, weather conditions often extended the time between observations, and additional downtime was required for telescope repairs and maintenance. Despite these obstacles, we ultimately achieved a median cadence of 3.9 days for Season 1. For Season 2, the median cadence for Field 1 was 5.5 days, 4.8 days for Field 2, and 5.6 days for Field 3.

\subsection{Template Images} \label{sec:templates}
We use images taken as part of the WHALES program as our template images. In the first season of our survey, all template images were captured during January 2023. We use the program SWarp to co-add images on a per-CCD basis \citep{swarp}. On average, each template results from the coaddition of 3 to 5 images with seeing less than 2.5 arcseconds and a 5 sigma magnitude limit \citep{1996A&AS..117..393B}. To observe SNe~Ia located beyond the Shapley Supercluster at a redshift of $z \sim 0.1$, where these SNe peak at $18.5$ mag, our templates reach a depth of $\sim$20th mag, ensuring that the templates do not contribute significant noise to the photometry. In Season 2, our templates were built from the images with the best seeing (less than 2.5\arcsec) from Season 1 and were similarly reduced.

\subsection{Classification} \label{sec:23m}
Most spectroscopic classifications for WHALES SNe come from the integral-field Wide-Field Spectrograph (WiFeS) mounted on the ANU 2.3m telescope at Siding Spring Observatory \citep{wifes1, wifes2}. Our spectroscopic selection process predominantly centers on identifying potential SN~Ia transients. Our current selection method relies heavily on visually inspecting sequences of images taken across multiple epochs of a likely candidate. We obtain spectroscopic confirmation of transients while the SN is live, giving higher priority to candidates near peak brightness, and no triggers have been too faint to follow up. However, some targets (approximately three a season) were classified as unknown or inconclusive, and due to observational constraints, we were unable to revisit them for further confirmation. External teams provided spectroscopic follow-up for four SNe across the two seasons, and these events are listed in Table~\ref{tab:SNe}.

\subsection{Redshifts} \label{sec: redshifts}
Redshifts for this sample were measured with the WiFeS spectrograph. Once the supernovae have faded, WiFeS was used to acquire host galaxy spectra. Host galaxies are identified using the directional light radius method (DLR; \citealp{sullivan_2006}), which links supernovae to their most probable hosts based on projected distance and luminosity. All targets for which redshifts were requested were successfully observed, and the redshifts obtained have a precision of $\sigma_z \sim 10^{-4}$ \citep{Hinton_2016}.

\subsection{Additional Data}
We supplement our data with data from the Asteroid Terrestrial-impact Last Alert System (ATLAS; \citealp[]{ATLAS}\footnote{\url{https://fallingstar-data.com/forcedphot/}}) project. which observes the night sky from Chile, Hawaii, and South Africa every two days in the optical. The observing bands are limited to the \textit{c} and \textit{o} bands (4200--6500~\AA~and 5600--8200~\AA, respectively). The survey reaches a depth of approximately $20$ mag and covers the full sky. ATLAS photometry is calibrated using Pan-STARRS \citep{Chambers16}, and its light curves are publicly available. While ATLAS discovers transients in the region of the Shapley Supercluster, the depth, typing, and completeness of discovery were uncertain when initiating the WHALES program. The WHALES survey complements ATLAS both as an additional discovery source and by providing more filter/color information as ATLAS employs two broader filters that span most of the optical window. This broader coverage is efficient for transient discovery, but the narrower SkyMapper bands provide more precise color information that is essential for light-curve fitting and distance calibration. Each SN discovered in the Shapley region was found by both surveys, improving confidence in the discovery completeness.

\section{SN Discovery and Light-curves } \label{sec: datared}
All images are initially processed through the SkyMapper real-time data reduction pipeline \citep{smpipe}. All CCDs in the focal plane are processed as a single unit for each exposure, where bad pixels are masked and corrections are applied for bias, flat-field illumination, overscan, and fringing ($i$-band only), as well as applying a WCS solution based on Gaia DR2 \citep{gaia_2018} that defines the transformation between pixel coordinates in an image and celestial coordinates. For the present analysis, the image subtraction components of the real-time pipeline are not used, allowing for customized post-processing steps tailored to this dataset.

The post-processing pipeline is based on the Corral Framework \citep{cor} and discussed in detail in Sections~\ref{sec:stelphot} and \ref{sec:sndisc}.

\subsection{Zeropoint Calibration}\label{sec:stelphot}

To establish zeropoints of the images, Point Spread Function (PSF) photometry is performed on sources in the science and template images using Source Extractor and PSFEx software \citep{1996A&AS..117..393B}. 

We use stellar sources from the SkyMapper DR4 catalog, which covers 21,000 square degrees of the sky \citep{smdr4}. These stars have been cross-matched with complementary surveys, including 2MASS \citep{Huchra} and Gaia \citep{gaia_2018}, to ensure precise astrometry and reliable photometry \citep{smdr4}. To enable consistent comparisons with other low-redshift SN~Ia datasets, we match SkyMapper stars to Pan-STARRS1 sources in the overlapping footprint based on Right Ascension (RA) and Declination (DEC), adopting a one-arcsecond matching tolerance. Each star is corrected for Milky Way extinction using IRSA\footnote{\url{https://irsa.ipac.caktech.edu/}}, following the implementation of \citet{Popovic2025} based on the \citet{Schlafly_2011} dust maps, with the correction applied at the effective wavelength of each filter.

To ensure linearity in the photometric comparison, we apply magnitude cuts following the approach of \citet{Brout_2022} and \citet{Popovic2025}, retaining only stars with $g > 14.8$~mag, $r > 14.9$~mag, $i > 15.1$~mag, and $z > 14.6$~mag. We further restrict the PS1 $g$-band sample to stars brighter than 19~mag to mitigate Malmquist bias, and limit the stellar color range to $0.25~\mathrm{mag} < g - i < 1.0~\mathrm{mag}$. We then compare the observed photometry to synthetic stars generated from the Next Generation Spectral Library (NGSL; \citealp{ngsl}), which provides model spectra for stars across a range of spectral types. This allows us to quantify systematic offsets in each band by fitting the observed magnitudes relative to the expectations from synthetic photometry. Using this method, we find magnitude offsets of $-0.0064 \pm 0.0003$ in $g$-band, $-0.0070 \pm 0.0003$ in $r$-band, and $-0.0106 \pm 0.0003$ in $i$-band. These offsets can be applied when combining this sample with other low-redshift SN compilations, such as Pantheon+ \citep{Scolnic_2022}.

\subsection{SN Discovery}\label{sec:sndisc}

\begin{figure}
    \centering 
    \includegraphics[width=\columnwidth]{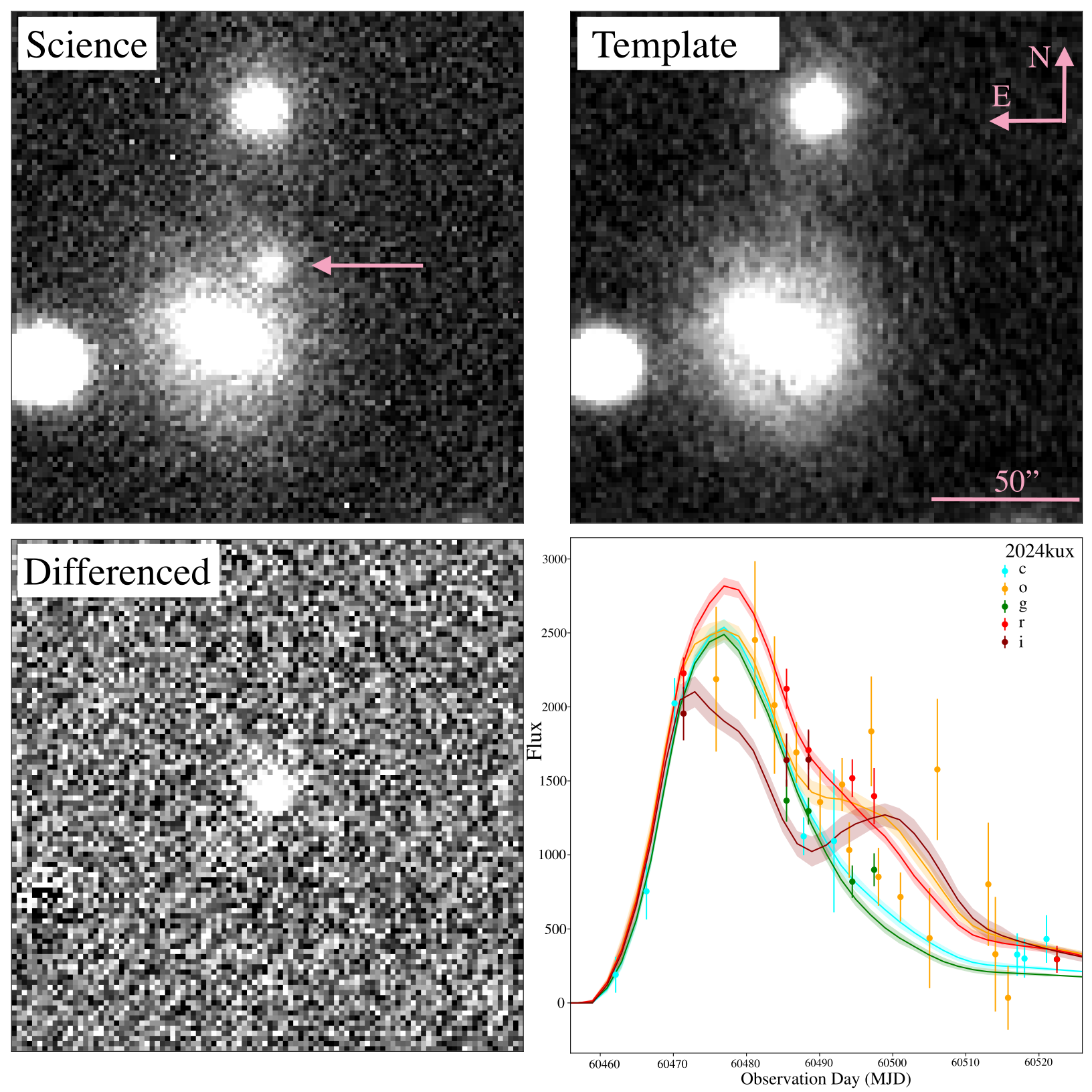}
    \caption{Examples image and light-curves from SN 2024kux. Images span $200'' \times 200''$ and are taken in the $r$-band. Upper Left: Science image from MJD 60471.4 (phase $-$5.5 days relative to optical maximum). Upper Right: Template image built from three images taken over a year before the explosion. Lower Left: Difference image of the upper two panels with the template image convolved to fit the science image. Lower Right: SNANA SALT3 SN~Ia fit to the light-curve.}
    \label{fig:diffim}
\end{figure}

To discover SNe, science and template images are astrometrically aligned. Difference imaging is then carried out using the image subtraction software High Order Transform of PSF And Template Subtraction (HOTPANTS; \citealp[]{hotp}), which utilizes the Alard \& Lupton algorithm \citep{Alard_1998}. The template image is convolved with three best-fit Gaussian kernels so that its PSF matches that of the science image. The template image is then scaled so that both images have the same zeropoint, and the template image is subtracted from the science image. An example of these images is shown in Figure~\ref{fig:diffim}. We also inject fake point sources, onto the CCD images as a quality control measure to ensure these objects are found by the pipeline.

We apply a set of selection criteria within a $10\times10$ pixel stamp around each detected object to filter real sources. We require a signal-to-noise ratio (S/N) greater than 5, that all pixels have positive, non-zero flux in both the science and template images, and that the source has a Source Extractor magnitude brighter than 23 and an elongation less than 2. The Source Extractor magnitude refers to the brightness of the source as measured by the Source Extractor software \citep{1996A&AS..117..393B}, which estimates magnitudes using either aperture or automated profile-fitting methods and a default zeropoint of 25 mag. The elongation is the ratio of the source’s semi-major axis to semi-minor axis, providing a measure of how stretched or circular the object appears; values near one correspond to roughly circular sources, while higher values indicate more elongated shapes. 

Objects meeting these criteria are referred to as ``valid detections.'' An ``associated candidate'' is defined when two or more detections exhibit measured positions within a 1.5-arcsecond proximity, regardless of the band or night they were observed. All associated candidates are saved, including objects that are not SNe~Ia, such as variable stars or fast transients. To identify the most promising SN~Ia candidates, we require that each object have more than three detections on separate nights, that it does not coincide with the position of any known star, that its maximum signal-to-noise ratio exceeds 5, and that it is detected in at least two separate filters. These criteria effectively filter out variable or spurious sources, leaving a set of “good associations” that are most likely to correspond to genuine supernovae. These good associations are then further filtered through a machine learning model trained on a subset of our images that have been manually labeled to distinguish good objects (either discovered SNe in our data or fake point sources) from others.

We identify 29 transient candidates with light curves potentially consistent with SNe~Ia before any spectroscopic confirmation across both seasons. Table~\ref{tab:SNe} offers a comprehensive overview of the transient sample observed during the first two seasons of the WHALES survey. It includes the transient name, classification type, sky location (RA and DEC), redshift ($z$), and the groups responsible for both discovery and spectroscopic classification according to the Transient Name Server. Although WHALES identified these transients independently, the discovery survey listed in Table~\ref{tab:SNe} reflects the group credited as the first to report the transient to the Transient Name Server (TNS). Because TNS assigns discovery credit to the earliest public report, WHALES does not appear in the table as the discovery survey, even for events we detected ourselves. In this sense, our role was in independent detection and follow-up, rather than being the first to publicly announce the transient. In several cases, we initiated requests for spectroscopic confirmation of our independently identified candidates. However, because the corresponding observations and subsequent reports to TNS were carried out under the Dark Energy Bedrock All-Sky Supernovae (DEBASS) program, the associated discovery survey recorded in Table~\ref{tab:SNe} is DEBASS \citep{nora, me}.

\begin{table*}[!t]
\centering
\caption{29 transients observed by WHALES in 2023 and 2024}\label{tab:SNe}
\begin{tabularx}{\linewidth}{l@{\extracolsep{\fill}}ccccccr}
Name & Type & RA [deg] & DEC [deg] & $z$ & $z_{err}$ & Disc.~Group & Class.~Group \\\hline\hline 
2023yk & -- & 202.694 & $-$32.093 & -- & -- & ATLAS & -- \\
2023dpc & SN~Ia & 204.973 & $-$30.932 & -- & 0.0001 & ATLAS & DEBASS \\
2023dpj & SN II & 201.435 & $-$29.837 & 0.0138 & 0.00015 & \citetalias{Hosenie_2021} & MeerLICHT \\
2023dtk & -- & 202.008 & $-$31.780 & 0.049007 & 0.0015 & ATLAS & -- \\
2023egt & SN~Ia & 201.451 & $-$30.711 & 0.045 & 0.0001 & ATLAS & DEBASS \\ 
2023egn & -- & 203.822 & $-$31.439 & 0.047337 & 0.0001 & ATLAS & -- \\
2023fvg & -- & 197.691 & $-$29.441 & -- & -- & ATLAS & -- \\
2023fqz & SN~Ia pec & 204.541 & $-$31.827 & 0.0398 & 0.0001 & ATLAS & DEBASS \\
2023jrb & -- & 202.895 & $-$31.812 & 0.1462 & 0.0015 & ATLAS & -- \\
2023adrv & -- & 203.787 & $-$30.119 & -- & -- & \citetalias{Groot_2024} & -- \\
2023oes & -- & 208.615 & $-$27.836 & 0.037009 & 0.0015 & ATLAS & -- \\
2024ayv & SN II & 207.090 & $-$29.628 & -- & -- & ATLAS & \citetalias{pessto} \\
2024cuw & -- & 201.267 & $-$31.06 & 0.05167 & 0.0015 & ATLAS & -- \\
2024cuv & -- & 201.034 & $-$31.486 & -- & -- & ATLAS & -- \\
2024ehb & -- & 211.162 & $-$33.320 & -- & -- & \citetalias{Hodgkin_2021} & -- \\
2024ecg & -- & 198.103 & $-$27.460 & -- & -- & ATLAS & -- \\
2024fee & SN~Iax[02cx-like] & 200.843 & $-$26.113 & 0.0314 & 0.00015 & GOTO & ePESSTO+ \\
2024fbo & SN~Ia & 196.555 & $-$30.161 & 0.011501 & 0.00015 & ATLAS & ePESSTO+ \\
2024fsb & -- & 199.303 & $-$27.526 & -- & -- & ATLAS & -- \\
2024fsd & SN II & 204.128 & $-$32.346 & -- & -- & ATLAS & DEBASS \\
2024hud & SN~Ia & 210.512 & $-$31.884 & 0.05 & 0.0001 & ATLAS & DEBASS \\
2024hue & SN~Ia & 211.527 & $-$31.558 & 0.068 & 0.0001 & ATLAS & DEBASS \\
2024jhc & SN~Ia & 196.648 & $-$29.732 & 0.075 & 0.0001 & ATLAS & DEBASS \\
2024kux & -- & 203.766 & $-$29.262 & 0.049494 & 0.0015 & ATLAS & -- \\
2024mtv & -- & 204.254 & $-$33.736 & -- & -- & BlackGEM & -- \\
2024ndk & -- & 198.570 & $-$31.719 & -- & -- & GOTO & -- \\
2024pft & SN~Ia & 198.146 & $-$29.966 & 0.074 & 0.0001 & ATLAS & DEBASS \\
2024pwn & SN~Ia & 207.937 & $-$29.337 & 0.047 & 0.0001 & ATLAS & DEBASS \\
2024sfd & SN~Ia & 199.024 & $-$25.690 & 0.0507 & 0.0001 & ATLAS & DEBASS \\ \hline
\end{tabularx}
\end{table*}

\subsection{First Light Curves}

With this final subset of associations, we perform forced PSF photometry at the precise location of the object in all overlapping images, regardless of whether a detection was initially flagged. This collection of forced photometry flux measurements and their associated uncertainties, acquired across multiple images, constitutes the light-curve dataset that will be used in our analysis. We show the light curves for this sample in Figure~\ref{fig:lc}.

We also include ATLAS data, as the WHALES and ATLAS surveys complement each other in both cadence and depth. By combining the early-time cadence of ATLAS with the additional bands of WHALES, we can construct well-sampled light curves that improve our ability to detect and classify SNe~Ia, and place them on the Hubble diagram with reduced uncertainties. 

\subsection{Repeatability of Stellar Magnitudes}

\begin{figure}
    \centering 
	\includegraphics[width=\columnwidth]{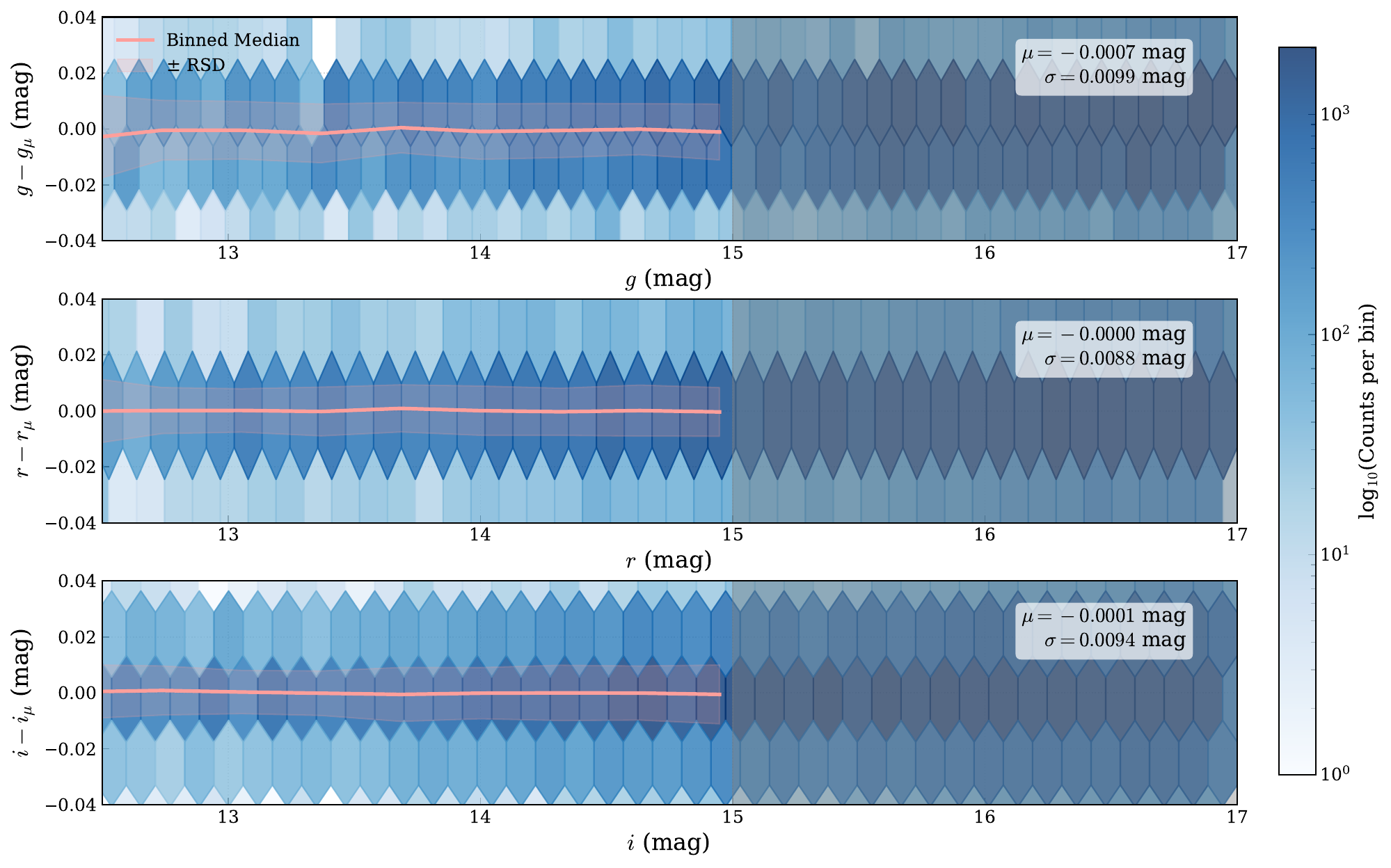} 
      \caption{{Evaluating the consistency of measured magnitudes for a given star by analyzing the distribution of observed values relative to the mean. Each magnitude is compared to the average and plotted as a function of magnitude and represented in blue based on the number of data points within each bin. The legend provides the overall median and RSD of the residuals for stars brighter than 15 mag, while binned statistics are shown in pink.}}
    \label{fig:sr1} \vspace{-1mm}
\end{figure}

We assess our photometric precision by examining the repeatability floor of the stars in our dataset. Following \citet{Peterson_2023}, we combine the stellar detections from all images for each SN. The mean magnitude of each star's magnitude distribution is computed and compared throughout the sample. Only stars found at the 15th magnitude or brighter are used. We acquire the robust median absolute standard deviation (RSD, \citealp[]{Hoaglin}; calculated using the median absolute deviation and scaled by a factor of 1.48) values of 0.0147, 0.0141, and 0.0141 mag for $g$, $r$, and $i$, respectively, as the scatter in Figure~\ref{fig:sr1}, and hence use 0.015 as our repeatability floor. Thus, we incorporate this as an extra error when calculating SN magnitudes and magnitude errors.

\section{First Data and Hubble Diagram points}\label{sec:datahubble}
\subsection{Light Curve Fits}
\begin{figure*}
    \includegraphics[width=\textwidth]{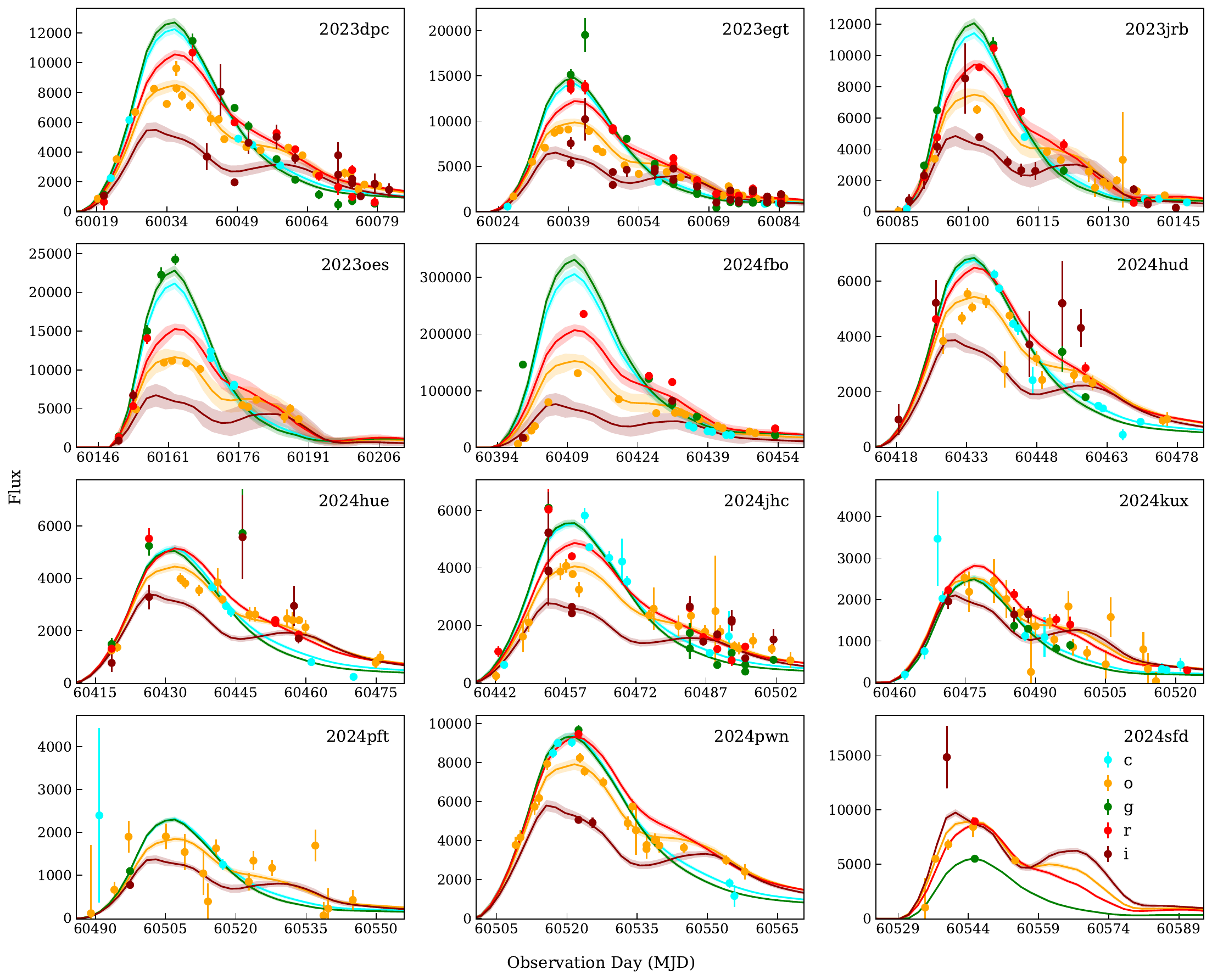}
    \caption{SALT3 light curves for the combined WHALES and ATLAS data.}
    \label{fig:lc}
\end{figure*}

We fit our light curves using the \texttt{SNANA} software package \citep{Kessler09, Kessler19} with the SALT3 model \citep{Kenworthy21}. When fitting using the SALT3 model, each SNe has a best-fit stretch ($x_1$), color ($c$), and amplitude ($x_0$), which are then used in a modified version of the Tripp equation \citep{Tripp98} to calculate the distance modulus, $\mu$,
\begin{equation}
    \mu = m_B + \alpha x_1 - \beta c - M,
\end{equation}
where $m_B$ is the apparent SN peak magnitude in the \textit{B}-band and is directly related to $x_0$ ($m_B = -2.5\log(x_0) + const$), $\alpha$ and $\beta$ specify the amplitude of the stretch-luminosity and color-luminosity corrections, and $M$ is the globally fit absolute SN peak magnitude. In this work, we set $\alpha = 0.14$ and $\beta = 3.1$ as found in \citet{Brout_2022}. The SALT3 fitted light curves for the WHALES SNe are shown in Figure~\ref{fig:lc}.

\begin{figure}
    \centering 
	\includegraphics[width=\columnwidth]{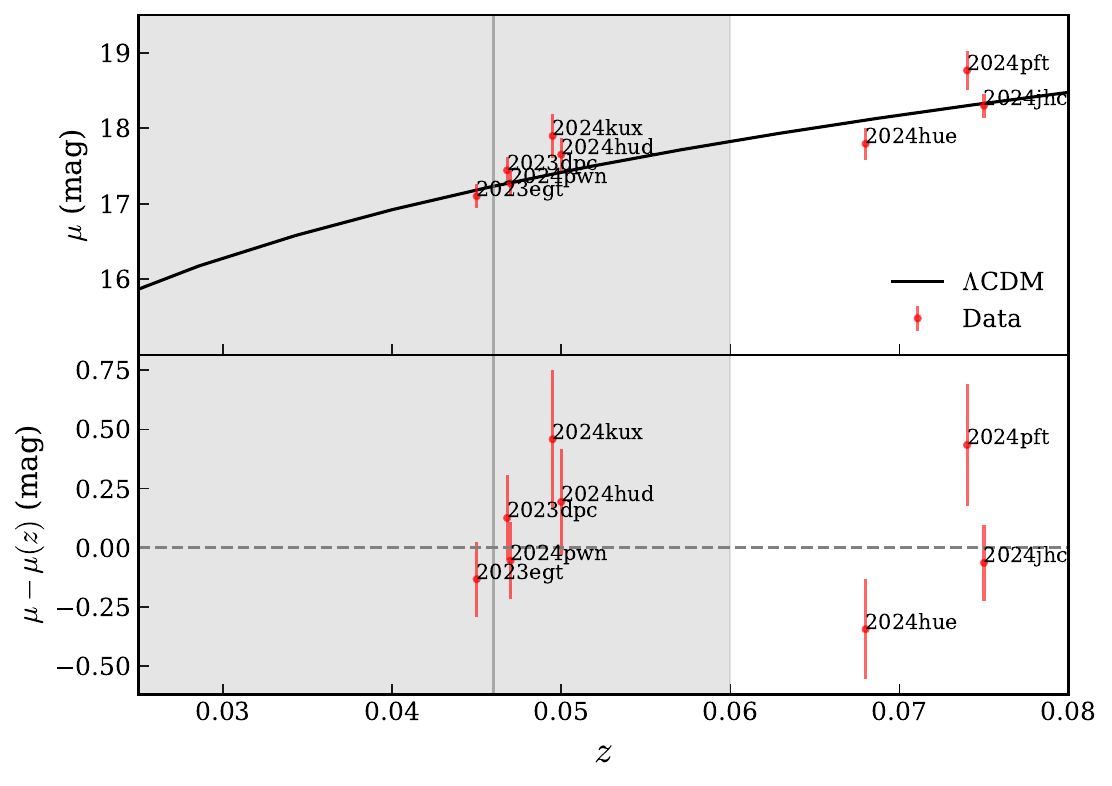} 
      \caption{Hubble diagram (top) and Hubble residuals (bottom). The shaded region represents the redshift range of the Shapley Supercluster as defined in \cite{Quintana_2020} ($0.02<z<0.06$) with the darker line marking the center redshift of the supercluster.}
    \label{fig:hd} \vspace{-1mm}
\end{figure}

We perform the following quality cuts on the WHALES data: SALT3 $|x_1| < 3$, $\sigma_{x_1} < 1$, SALT3 $|c| < 0.3$, and $\sigma_{t_{peak}}<2$ days. In total, 8 out of 12 SNe~Ia satisfy all of these cuts; 2023jrb, 2023oes, 2024fbo, and 2024sfd all fail the color cut (with $c$ of $-$0.342, $-$0.571, $-$0.723, and 0.674 respectively). The $\mu$ values without peculiar velocity corrections are shown in a Hubble Diagram in Figure~\ref{fig:hd} along with the Hubble residuals for a $\Lambda$CDM cosmology. We discuss how to interpret this Hubble diagram in the following section.

\section{Discussion}\label{sec:disc}

\subsection{Simulations of the Impact of A Supercluster on the Distance-Redshift Relation}

To assess the accuracy of measuring motions around superclusters, we replicate SNe~Ia in the Shapley Supercluster using the Uchuu cosmological simulation \citep{Ishiyama_2021, Aung_2022}, adopting the same friends-of-friends (FoF) grouping schema described in \citet{Peterson_2025} (see their section 3.2.1). In brief, \citet{Peterson_2025} employ a modified FoF algorithm, originally developed for the 2MRS galaxy-group catalog \citep{Lambert_2020}, that is rooted in graph theory and tailored for observational data. 

To identify superclusters, we define groups of galaxy clusters as belonging to the same structure if they are separated by less than $12~h^{-1}~\mathrm{Mpc}$. This separation was determined empirically from the volume of the Shapley Supercluster (expressed in $10^{8}h^{-3}\mathrm{Mpc}^3$) and the number of galaxies it contains ($\sim10^4$). For each realization, we place an observer at a random location on a sphere around the supercluster such that the distance to the center matches that of our distance to Shapley.

\begin{figure}
    \centering
    \includegraphics[width=\columnwidth]{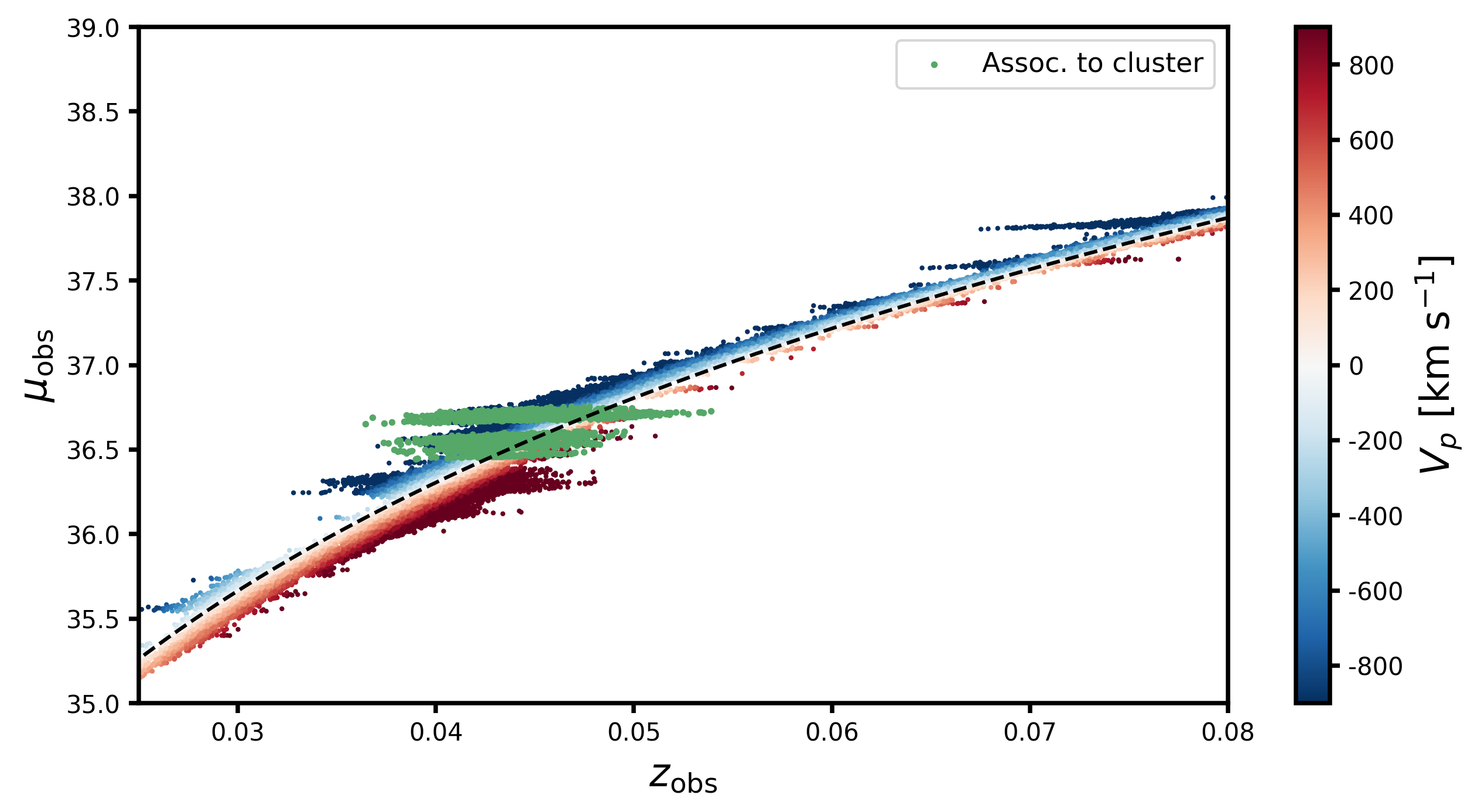}
    \includegraphics[width=\columnwidth]{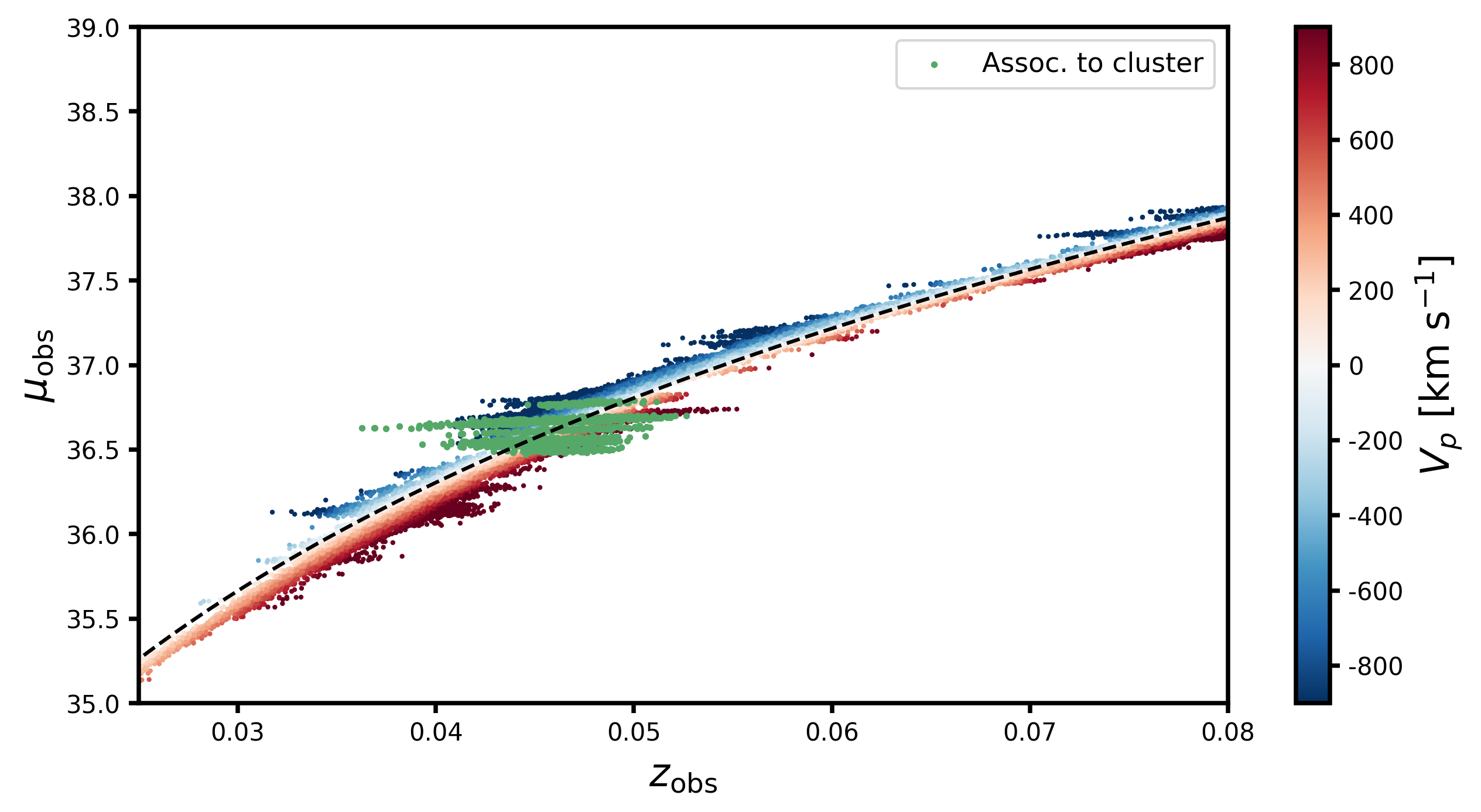}
    \caption{Simulated Hubble diagrams for SNe in the vicinity of Shapley-like superclusters with no SN~Ia scatter for two different realizations of a supercluster. Each point corresponds to a SN and is color-coded by peculiar velocity ($v_{p}$), with red (positive) indicating galaxies moving away faster than the Hubble flow and blue (negative) indicating infall. SNe associated with the central cluster are highlighted in green. }
    \label{fig:simulations}
\end{figure}

Our simulations highlight the diversity of possible realizations for a Shapley-like supercluster. Figure~\ref{fig:simulations} shows two representative cases: the first containing $\sim$1400 galaxies hosting SNe with a total mass of $2.84\times10^{15} M_{\odot}$, and the second with $\sim$2200 galaxies and a mass of $5.36\times10^{15} M_{\odot}$. In both realizations, the internal motions of galaxies within their host halos produce a pronounced elongation along the line of sight, a well-known phenomenon known as the Fingers-of-God effect \citep{Jackson1972}. This distortion complicates attempts to disentangle subclusters from the large-scale velocity field, underscoring the importance of precise distance indicators such as SNe~Ia. We note that in these simulations, no intrinsic scatter has been introduced for the SNe.

The challenges we identify for Shapley are reminiscent of those encountered in recent efforts to constrain the mass of the Coma Cluster \citep{Benisty_2025}. In Coma, the central region is effectively virialized, allowing its dynamics to be modeled robustly; however, the situation becomes more complex when multiple such clusters interact within a larger supercluster. In this context, the Shapley Supercluster can potentially be regarded as an $N$-cluster problem, where the global dynamics reflect the combined and overlapping gravitational influences of several virialized subclusters. Modeling such a system requires extending beyond single-cluster mass reconstruction methods and represents a promising avenue for future work.

In Figure~\ref{fig:hd}, our Hubble diagram does not show the characteristic fingers in redshift space, likely as a consequence of the limited statistics in our current sample. Recovering such a substructure will be challenging even with ideal data: as our simulations show (Figure~\ref{fig:simulations}), the expected separation between Shapley’s subclusters corresponds to only $\sim$0.1~mag in distance modulus. While our current sample is too small to reveal a clear substructure, with a sufficiently large SN~Ia sample, the Hubble diagram itself can serve as a kinematic probe of cluster and supercluster mass. 

\subsection{Supernova Rates in a Supercluster}

In Season 2, when we observed the full Shapley Supercluster, we found 8 SNe~Ia. To understand if this is the expected amount, we can calculate the rate of SNe~Ia in two ways. The first way uses the volumetric rate of explosions measured by the Palomar Transient Factory (PTF), as presented in \cite{Frohmaier_2019}:

\begin{multline}
    r_{v,\mathrm{F19}}(z) = (2.27 \pm 0.19) \times 10^{-5} \ \mathrm{Mpc}^{-3} \ \mathrm{yr}^{-1} \ 
    \\ \times (1+z)^{1.70 \pm 0.21}.
\end{multline}

We rescale this rate to our fiducial value of the Hubble constant using

\begin{equation}\label{eq: sn_rate}
r_v = r_{v,\mathrm{F19}}\left(\frac{h}{0.7}\right)^3.
\end{equation}

Such rates, derived from wide-field surveys, provide a census of SN~Ia occurrences per unit volume as a function of redshift and thus serve as a natural starting point for forecasting detections. We compute the expected number of SNe~Ia in the redshift range $0.01 < z < 0.08$ and for a survey area of $200\ \textrm{deg}^2$, over a 6-month observing season. The upper redshift limit of 0.08 is calculated to be near the maximum SNe redshift discovered as part of this program ($z=0.075$). Since we cover approximately 200 square degrees every three nights, we assume that we are close to 100\% efficient in detecting SNe~Ia within this region. Assuming a Poisson distribution, this yields an average of $9 \pm 3$ SNe~Ia per season. 

While such a volumetric calculation provides a broad statistical baseline, it does not account for the distinctive environment our survey probes. A separate path is to calculate the rate using the number of galaxies in the Supercluster using a mass-dependent rate estimate. Assuming a population of 10,000 galaxies with an average stellar mass of $10^{10} M_{\odot}$, the total stellar mass is $10^{14} M_{\odot}$. Using a specific SN~Ia rate of
\begin{equation}
r_{\mathrm{Ia}} = 0.2 \ \mathrm{SNe} \ (10^{10} \ M_{\odot})^{-1} \ \mathrm{century}^{-1}.
\end{equation}
\noindent as defined in \citet{toy_2023} (see their figure 8), dividing the total stellar mass by $10^{10} M_{\odot}$ and applying the specific rate of $0.2$ SNe~Ia per $10^{10} M_{\odot}$ per century yields an expected rate of $\sim$2000 SNe~Ia per century or 10 per observing period. An important subtlety is that our assumed mass of $10^{14} M_\odot$ represents only the stellar mass in galaxies, not the total mass. For comparison, \citet{stopyra2021} report a total mass for the richest Shapley subcluster (A3571) in the range $(3 \times 10^{14} - 2 \times 10^{15}) M_\odot/h$, meaning our adopted value corresponds to roughly 5–30\% of that range. It is also far smaller than the $8.9 \times 10^{16}\ h^{-1}\ M_\odot$ total mass of Shapley measured in the 2M++ galaxy redshift catalog, which includes both dark matter and gas, and corresponds to a galaxy and stellar mass density approximately 5–10 times higher than the cosmic mean across its core region \citep{Einasto_2007, Lavaux_2011}. In general, galaxies are expected to contribute only about 2–5\% of a supercluster’s total mass, though whether our adopted stellar mass estimate is fully consistent with this fraction remains uncertain \citep{mass_2010}. 

The two independent estimates yield nearly the same expectation value, of order $\sim$10 SNe~Ia per observing season. The volumetric method assumes a homogeneous and isotropic distribution across the redshift interval $0.01 < z < 0.08$, while Shapley itself occupies a narrower slice of this range. The convergence between the two estimates arises because, theoretically, on sufficiently large scales, overdense regions such as Shapley are statistically compensated by neighboring underdensities. The volumetric rate effectively averages over both, resulting in a number that matches the direct mass-based estimate. Measurements of the galaxy angular correlation function support this interpretation and, as shown by \citet{martin_2011}, the correlation amplitude becomes negative on scales of $\theta \sim7–9^{\circ}$ (their figure~3, top panel), corresponding to the angular extent of the Shapley Supercluster. A negative correlation at this scale indicates that overdensities of Shapley’s size are theoretically balanced by surrounding voids, consistent with the expectation from large-scale homogeneity. The similarity between the volumetric and mass-based estimates therefore reflects not coincidence, but the statistical averaging of overdensities and underdensities in a homogeneous and isotropic universe.

Taken together, these calculations suggest that our estimated rate of $\sim$10 SNe~Ia per observing season is reasonable. This implies that building statistically powerful samples in Shapley will require sustained monitoring over multiple seasons and the contribution of future wide-field surveys such as LSST. Expanding the SN~Ia sample with these precision requirements will be essential for obtaining robust cluster and supercluster mass constraints.

To estimate the precision with which SNe~Ia can constrain the mass of the Shapley Supercluster, we perform a Fisher forecast following the formalism of the Hubble flow model presented by \citet{Benisty_2025}. Specifically, we follow equation 23 in \citet{Benisty_2025} which presents the velocity-distance relationship in terms of the mass of the cluster and $H_0$. In our framework, the peculiar velocities of SNe~Ia are modeled as tracers of the underlying gravitational potential, allowing the total mass of a structure to be inferred from deviations in the Hubble flow. We restrict our analysis to SNe beyond 10~Mpc/h from the cluster center to exclude objects dominated by random virial motions, which do not contribute useful information about the coherent infall pattern. Assuming a total mass of $10^{16}M_\odot$, representative of the Shapley Supercluster, we find that a sample of 100 well-measured SNe~Ia yields a forecasted precision of better than 25\% on the total mass.

\section{Conclusion}\label{sec:concl}
This work represents the first supernova survey to deliberately target a supercluster as its discovery region. Over two observing seasons, WHALES monitored $\sim$200 deg$^2$ of the Shapley Supercluster and identified 12 SNe~Ia, 8 of which are spectroscopically confirmed. These discoveries demonstrate the feasibility of building a dedicated supernova sample in one of the most massive structures in the local universe. To interpret these observations, we turned to simulations, which revealed the importance of accounting for the internal structure of Shapley: its population of subclusters introduces distinctions that cannot be ignored in any robust mass estimate. A full treatment of this complexity lies beyond the scope of the present study, but the path forward is clear. Building a dedicated sample of SNe~Ia within this region is essential for distinguishing between the subclusters.

Our simulations demonstrate that a distance precision of order 0.1 mag is required to separate these structures, a level of accuracy that SNe~Ia are uniquely suited to provide. With the survey data already in hand, and with the prospect of future contributions from large-scale programs such as LSST, the available sample of supernovae in Shapley will continue to grow. As this sample increases, it will enable the level of precision needed to disentangle the substructure of the supercluster and to apply mass estimation techniques such as those outlined by \citet{Benisty_2025}.

In summary, while the present study establishes only the groundwork, it highlights the crucial role of SNe~Ia in bridging observations of Shapley with the theoretical frameworks required for precise mass estimates. The supernova sample thus provides the necessary foundation for future efforts to measure not only the global mass of the Shapley Supercluster, but also the contributions of its constituent subclusters.

\section{Software and third party data repository citations} \label{sec:cite}
astropy \citep{2013A&A...558A..33A,2018AJ....156..123A,astropy:2022}, matplotlib \citep{Hunter07}, numpy \citep{numpy11}, \texttt{pandas} \citep{the_pandas_development_team_pandas-devpandas_2024}, \texttt{PIPPIN} \citep{Hinton2020}, scipy \citep{scipy}, seaborn \citep{Waskom2021}, \texttt{SNANA} \citep{snana}.

\begin{acknowledgments}
We thank the Templeton Foundation for directly supporting this research (D.~S.). D.~S. is supported by Department of Energy grant DE-SC0010007, the David and Lucile Packard Foundation, and the Templeton Foundation. D.~S. and M.~A. are supported by the Alfred P. Sloan Foundation. This material is based upon work supported by the National Science Foundation Graduate Research Fellowship under Grant No. DGE 2139754. We thank the reader. 
\end{acknowledgments}

\begin{contribution}

M.~A. performed the analysis and wrote the majority of the paper. D.~S. advised and provided critical review and editing. B.~C. contributed to the simulation aspects of the analysis and participated in the review and writing. E.~R.~P. contributed to the light curve fitting and the review and formatting of the paper. B.~O.~S contributed to the survey observing strategy and the discovery analysis pipeline software. All authors provided editing and discussions around the paper.


\end{contribution}

%
\facilities{SkyMapper, ANU 2.3m}

\software{
          astropy \citep{2013A&A...558A..33A,2018AJ....156..123A,astropy:2022}, 
          matplotlib \citep{Hunter07}, 
          numpy \citep{numpy11}, 
          \texttt{pandas} \citep{the_pandas_development_team_pandas-devpandas_2024}, 
          \texttt{PIPPIN} \citep{Hinton2020}, 
          scipy \citep{scipy}, 
          seaborn \citep{Waskom2021}, 
          \texttt{SNANA} \citep{snana}
          }


\bibliography{sample701}{}
\bibliographystyle{aasjournalv7}



\end{document}